# Search for environmental effects on the KLL Auger spectrum of rubidium generated in radioactive decay

A.Kh. Inoyatov[a,b], L.L. Perevoshchikov [a], A. Kovalík[a,c], D.V. Filosofov[a], Yu.V. Yushkevich[a],
M. Ryšavý[c], B.Q. Lee[d], T. Kibédi[d], A.E. Stuchbery[d], V.S Zhdanov[e],

[a] *Laboratory of Nuclear Problems, JINR, Dubna, Moscow Region, Russian Federation*
[b] *Institute of Applied Physics, National University, Tashkent, Republic of Uzbekistan*
[c] *Nuclear Physics Institute of the ASCR, CZ-25068 Řež near Prague, Czech Republic*
[d] *Department of Nuclear Physics, RSPE, The Australian National University,
   Canberra, ACT 0200, Australia*
[e] *Nuclear Physics Institute, Almaty, Kazakhstan*

*Keywords:* Electron spectroscopy; Ion implantation; Vacuum evaporation; $^{83}$Rb; $^{85}$Rb; $^{83}$Sr; $^{85}$Sr; Auger effect; KLL transitions; Transition energy; Transition intensity; Chemical shift; Atomic environment; Atomic structure effect; Natural width of Auger line; Natural atomic level width; Multiconfiguration Dirac-Fock calculations;

**Abstract**

The KLL Auger spectrum of rubidium following the electron capture decay of $^{83}$Sr and $^{85}$Sr isotopes was experimentally studied in detail for the first time using one $^{83}$Sr source and three $^{85}$Sr sources in different host matrices. Energies, relative intensities, and natural widths of all the nine well-resolved basic spectrum components were determined and compared with both predictions and experimental data for krypton. Results of our multiconfiguration Dirac-Fock calculations demonstrated an influence of the "atomic structure effect" on absolute energies of the KLL transitions following the creation of initial vacancies by the electron capture decay. Environmental effects on the KLL Auger spectrum were distinctly observed only for the absolute transition energies.

## 1. Introduction

The KLL Auger spectrum is the most intense and the simplest among the K Auger groups, having only nine spectrum components. At the present time, the basic structure of the spectrum for elements across a wide atomic number (Z) range can be satisfactorily described by, e.g., the semi-empirical transition energy calculations [1] and the relativistic transition rate calculations [2,3]. These calculations [1-3] were performed in the intermediate coupling framework and took into account the configuration interaction.

In the last two decades, the predicted strong influence [3] of the relativistic effects on the intensity distribution between the $^{3}P_{0}$ and $^{1}P_{1}$ components of the $KL_{1}L_{2}$ doublet have been experimentally proved (see, e.g., [4] and references therein) for several elements in a relatively wide Z interval (see Fig. 1). It is generally known that the lower Z, the smaller influence of the relativistic effects can be expected. However, it is still an open question of up to which Z this influence can be neglected.

In some previous experimental works (see, e.g., [4] and references therein), the basic theoretical assumption that the Auger process following the electron capture (EC) decay is independent of the prehistory of the primary vacancy was questioned. But the detailed experimental investigation is complicated by presence of some other effects like the influence of the physicochemical atomic environment etc.

There is also a lack of experimental data on the influence of atomic environments on the KLL Auger spectra especially for the medium and heavy elements. But this information is also indispensable for the interpretation of weak effects in experimental Auger electron spectra as mentioned above.

In the present paper we report results of our experimental investigation of the KLL Auger spectrum of rubidium following the EC decay of the $^{83}$Sr ($T_{1/2}$=32.4 h) and $^{85}$Sr ($T_{1/2}$=64.9 d) isotopes incorporated into different host matrices in order to address the stated physical problems. The $^{85}$Sr sources used were prepared by vacuum evaporation on a polycrystalline carbon backing as well as by ion implantation at 30 keV into both a high purity polycrystalline platinum foil and the mentioned carbon substrate. Likewise, the source of $^{83}$Sr was prepared by ion implantation at 30 keV into the same Pt foil.

Our investigations were performed in the frame of the development of a new technique for the preparation of super stable calibration $^{83}$Rb/$^{83m}$Kr electron sources [6,7] for the KATRIN neutrino mass experiment [8]. It should be noted that the KLL Auger spectrum of rubidium (Z=37) was experimentally studied for the first time in this work.

## 2. Experimental

### 2.1. Source preparation

Radioactive isotopes of strontium were obtained by spallation of metallic yttrium by 300 MeV protons from the internal beam of the phasotron particle accelerator at the JINR, Dubna, Russia. After the "cooling" for three days, the irradiated target (1 g weight) was dissolved in concentrated nitric acid. Strontium isotopes were chemically separated from the target material and other elements using "Sr resin" (TrisKem International). An additional purification was then carried out on a cation-exchange chromatography column (70 mm length, 2 mm diameter, A6 resin) also in a nitric acid medium. Afterwards the strontium fraction ($^{82}$Sr ($T_{1/2}$=25.3 d), $^{83}$Sr ($T_{1/2}$=32.4 h), $^{85}$Sr ($T_{1/2}$=64.9 d)) obtained was used for both methods of electron source preparation, namely mass separation and vacuum evaporation.

The mass separation of the strontium isotopes was performed on a mass separator at the JINR, Dubna. At the same time, the strontium ions were embedded (at the energy of 30 keV) into the platinum or carbon foils. Prior to use, the surface of the Pt foils was cleaned by alcohol. Parts of the foils containing the Sr isotopes with the atomic mass number A = 83 and 85 were cut out and used for the electron spectrum measurements. The typical size of the "active" spots was about 2x2 cm$^2$. The activity of the $^{85}$Sr isotope in the sources upon the preparation was 950 and 380 kBq for the platinum and carbon backings, respectively, and 11.7 MBq of $^{83}$Sr in the corresponding source prepared.

According to the simulations [6] performed for the implantation of $^{83}$Rb ions at 30 keV into the same type of the high purity polycrystalline platinum foil as used in our experiment, the mean projected range and standard deviation of the distribution of the implanted atoms along the depth of the foil amounted to about 9 and 4 nm, respectively. Real circumstances of the implantations [6] such as the zero ion incident angle (relative to the source foil normal), polycrystalline structure of the Pt foil, and an adsorbed surface contamination layer of water and hydrocarbons (so called "rest gas layer") were represented in the simulations by an additional 3 nm thick pure carbon layer on the foil surface. These experimental conditions were very close to our ones. It was, moreover, found in Ref. [6] that a portion of 5-10% of the incident $^{83}$Rb ions gets stopped in the "rest gas layer", which represents a non-metallic environment. As surfaces of the Pt foils were not cleaned (by ion sputtering or any other means) after the $^{83}$Rb ions collections, a certain portion of $^{83}$Rb thus remains in this non-metallic environment. This feature pertains to our $^{83}$Sr and $^{85}$Sr sources. After the collections of the $^{83}$Sr and $^{85}$Sr ions, the sources were exposed to air during transfer to the electron spectrometer. Their surfaces were also not cleaned before the electron spectra measurements. Thus a portion of Sr ions stopped on the Pt foil surfaces in the above non-metallic environment were bound with atoms of oxygen in all possible forms (oxides, hydroxides, carbonates, hydrocarbonates, etc.) and had the oxidation number +2 (see below).

Thermal evaporation deposition on a polycrystalline carbon foil of 150 μm thickness took place at 1400 °C. Prior to use, the surface of the foil was mechanically cleaned. The strontium fraction was transferred to an annealed Ta evaporation boat and dried up. To remove possible volatile organic compounds, the Ta evaporation boat with deposited activity was first heated at 800 °C for about 30 s. The source backing was shielded throughout this procedure. During the evaporation, the source backing rotated around its axis at a speed of 3000 turns/min at a distance of 8 mm from the Ta evaporation boat to improve homogeneity of the evaporated layer. Altogether two different sources were prepared with the $^{85}$Sr activity of 2.3 and 1.1 MBq upon preparation.

The exact chemical state of the deposited trace amounts of $^{85}$Sr on the surfaces of the source backings in vacuum was unknown. However, the sources were transferred to the electron spectrometer in air after the preparation and thus the evaporated layers were exposed to it. Due to extreme strontium reactivity to air, ions of $^{85}$Sr were bound with atoms of oxygen in all possible forms (oxides, hydroxides, carbonates, hydrocarbonates, etc. of different proportions) and had the oxidation numbers +2. Nevertheless, the overwhelming majority of the parent $^{85}$Sr atoms were most likely in the $SrCO_3$ chemical form. This statement is based on: (i) specific chemical properties of strontium, (ii) its known macro-chemistry, (iii) the inner self-consistency of the physicochemical methods used for the preparation of the sources, and (iv) the conditions of their treatment. After the $^{85}$Sr EC decay, the daughter $^{85}$Rb atoms were stabilized in the above $^{85}$Sr matrices. The $^{85}$Rb ions thus were most likely bound with oxygen atoms in anions of all possible relevant forms ($O^{2-}$, $OH^-$, $CO_3^{2-}$, $HCO_3^-$, etc.). Contrary to $^{85}$Sr, they had the oxidation number +1.

*2.2. Measurements and energy calibration*

Electron spectra were measured in sweeps using a combined electrostatic electron spectrometer [9] consisting of a retarding sphere followed by a double-pass cylindrical mirror energy analyzer. The choice of the absolute spectrometer energy resolution and the spectrum scanning step depended on the intensity of the radioactive source being measured and the complexity of the inspected spectrum. Examples of the measured spectra are shown in Figs. 2-5.

Altogether 14 low energy conversion electron lines (listed in parentheses) of nuclear transitions in $^{169}$Tm with energies $E_\gamma$ = 8.41008(21) [11] ($M_{1,2}$, $N_{1,3}$), 20.74378(10) [12] ($L_{1-3}$, $M_{1-3}$, $N_{1-3}$), and 63.12081(5) keV [13] (K) along with five lines (K, $L_{1-3}$, $M_1$) of the 14.41300(15) keV [13] nuclear transition in $^{57}$Fe were used for calibration of the spectrometer energy scale. Energies of the calibration lines related to the Fermi level $E_F(i)$ ($i$ is the atomic subshell index) were evaluated as $E_F(i) = E_\gamma - E_{bF}(i)$ making use of the experimental Fe and Tm electron binding energies $E_{bF}(i)$ [14] related to the Fermi level. Experimental uncertainties from 0.4 to 1.6 eV and from 0.4 to 0.9 eV are quoted in Ref. [14] for thulium and iron electron binding energies, respectively. Thus in all cases the $E_{bF}(i)$ uncertainties fully dominate uncertainties of the above nuclear transition energies. Therefore uncertainties in the calibration line energies were almost identical with those of the relevant electron binding energies.

In the case of Fe, the electron binding energies [14] are given for an oxide but the chemical form for Tm is not specified. Chemical techniques applied in the preparation of our calibration sources by vacuum evaporation on a polycrystalline carbon substrate should guarantee the oxide state for the both elements. For the case this is not true, we took into account the maximum measured chemical shifts of the outer subshell electron binding energies of about 2 and 4 eV [15] among different chemical states of Tm and Fe, respectively. We found that the influence of these shifts on the energy calibration is well below the standard deviations of the measured energies of the studied electron lines (quoted in Table 1 and in the text).

*2.3. Spectra evaluation*

The measured electron spectra were evaluated using the approach and the computer code described in Ref. [16]. The individual spectrum line shape was expressed by a convolution of a Gaussian ("spectrometer response function to monoenergetic electrons") and an "artificial" function which describes the natural energy distribution of the investigated electrons leaving atoms (Lorentzian) distorted by manifold inelastic scattering of the electrons in the source material (surface and volume plasmon excitations, shake-up/-off, atomic and lattice excitations, etc.). In order to find the proper form of the complex discrete energy loss peak and its long low energy tail (going down to the "zero" energy), multiple fitting of the measured electron spectrum with random variations (Monte Carlo method) of the shape of the fitted lines in this "energy loss region" within the preset shape limits was applied. It should be noted, that the position and width of the discrete energy loss peak depend on the source material, while its intensity relies on a ratio of the energy dependent mean free path for inelastic electron scattering and the effective source thickness. (Contrary to the inelastically scattered electrons, the electrons which left the electron source without any energy loss create a so-called zero loss (or no-loss) peak. This peak can be described by a simple convolution of a Gaussian and a Lorentzian resulting in a Voigt function.)

In the evaluation, the fitted parameters were the position, the height, and the width (of the no-loss peak) of each spectrum line, the constant background and the width of the spectrometer response function. Results of the evaluations are shown graphically in Figs. 2, 3 (continuous lines) and numerically in Tables 1-4 and in the text. The quoted uncertainties are our estimates of standard deviations ($\sigma$).

## 3. Results and discussion

As can be seen from Fig. 2 and 3, all the nine basic spectrum components of the KLL Auger spectrum of Rb were clearly resolved in our measurements. It is also seen, that discrete energy loss peaks of the spectrum lines are much higher for the $^{85}$Sr sources prepared by vacuum evaporation on the carbon foil than for those prepared by ion implantation into the platinum backing. On the other hand, the low energy tails of the spectrum lines converge faster to the background level for the evaporated sources. This fact is caused by manifold scattering of the emitted electrons in the Pt foil.

In Tables 1 and 3, energies and relative intensities, respectively, of the basic KLL Auger lines for the two sources are compared with results of the transition energy calculations of Ref. [1] and this work and the relativistic transition rate calculations of both Ref. [3] and the present work. (Results of the works [3] and [2] are almost identical.) Changes in the absolute KLL transition energies induced by environmental effects are presented also in Table 2. In Table 4, the measured natural widths of the KLL Auger lines for the two different sources are displayed together with estimated values based on four sets of Rb natural atomic level widths listed in Table 5.

### 3.1. Transition energies

It is seen from Table 1 (the fourth and fifth columns) that the measured KLL transition energies in rubidium ("adopted values") relative to that of the most intense $KL_2L_3(^1D_2)$ spectrum line (i.e. the mutual energy line positions) are systematically greater than values obtained from the widely used semi-empirical calculations [1] (see the sixth column). The energy interval occupied by the calculated spectrum (from the first to the last spectrum line) is shorter by 5.5(4) eV than the experimental one (515.9 eV contrary to 521.4(4) eV), i.e. it is "compressed". The same situation was observed [17] for the nearest neighbor element Kr (Z=36) (see the seventh column of the table). In this case the theoretical spectrum [1] is shorter by 5.8(7) eV and differences between the experimental and calculated [1] relative transition energies are almost identical with those observed for rubidium. It seems improbable, therefore, that the discrepancies between the calculated [1] and measured relative KLL transition energies for rubidium (and also for Kr [17]) are caused by inaccuracies in the experimental electron binding energies used in the calculations [1]. They are likely caused by some imperfections of the physical model applied in the calculations [1].

The absolute energy (related to the Fermi level) of the dominant $KL_2L_3(^1D_2)$ transition in rubidium measured with the $^{85}$Sr source prepared by vacuum evaporation on the carbon foil is higher by 6.1(16) eV than the semi-empirical prediction [1] (see the second and the fifth columns), i.e. by almost 4 σ. Nevertheless the declared precision of the calculations [1] is generally 1-2 eV. The evaporated source should ensure the closest conditions for rubidium atoms to those under which the calculations [1] were performed. As can be seen from Tables 1 and 2, and the following discussion, a substantial change of the physicochemical environment of the $^{85}$Rb atoms brought a decrease of the $KL_2L_3(^1D_2)$ transition energy by about 3 eV, i.e. it works on the transition energy in the opposite direction. As mentioned in Ref. [18], this discrepancy may have a contribution from inaccuracies in some corrections (e.g. for the solid state) and approximations used in the semi-empirical calculations [1]. Nevertheless, it seems that the main cause of the increase in the experimental $KL_2L_3(^1D_2)$ transition energy stems from the so-called "atomic structure effect" (see, e.g., [19]) which was revealed for the first time in K X-rays of holmium. Higher energies for K Auger transitions following the EC decay can be explained as a result of additional screening of the daughter nucleus by a "spectator" electron because the $10^{-16}$-$10^{-17}$ s lifetime of the 1$s$ atomic hole produced in the EC decay is so short that the intermediate state has an outer-electron configuration close to that of the parent atom. In X-rays, the effect is the most pronounced for rare earth elements and especially for those from the 4$f$ and 5$f$ groups.

We therefore performed calculations of the KLL Auger transition energies in rubidium (Z=37) following the creation of initial vacancies by both the electron capture (EC) decay of $^{85}$Sr and the internal conversion (IC) processes in the $^{85m}$Rb daughter decays. In the calculations, the atomic configuration for the IC decay was assumed to be that of a neutral rubidium system (i.e. only one electron on the 5s shell), while for the EC decay, an extra 5s electron was included, corresponding to the 5$s^2$ valence-shell configuration of a neutral Sr atom. These "*ab initio*" multiconfiguration Dirac-Fock (MCDF) calculations used the actual electronic configurations of both the initial and final states in the KLL Auger transitions, including the vacancies. Both relativistic effects and quantum-electrodynamic (QED) corrections were taken into account. Each of the Auger transition energies were evaluated as the difference of the total energies of the initial and final states and are referenced to the vacuum level. The calculations were performed for vapor, i.e., without accounting for the solid state effects. A more detailed description of the calculations is given in in the Appendix of the paper.

As can be seen from Table 1 (the next-to-last column), our calculated energy for the $KL_2L_3(^1D_2)$ transitions following the EC decay (11441.8 eV) matches the measured value 11441.5(16) eV for the evaporated $^{85}$Sr source while that one (11434.8 eV) obtained for the $KL_2L_3(^1D_2)$ transitions following the internal conversion is lower by 6.7(16) eV and matches the semi-empirical prediction of 11435.4 eV [1] also calculated for the IC processes. These "agreements" are, however, seeming because the experimental and semi-empirical [1] transition energies are referenced to the Fermi level while our calculated values are related to the vacuum level and are, moreover, valid for vapor systems (as emphasized above). The Fermi and vacuum levels differ by the corresponding work function which amounts to about 5.7 eV for our spectrometer. If one uses this work function and follows the procedure [1] for correction of the transition energies for the solid state effect (increasing them) with a strontium solid state correction of 6.0 eV [1] (the matrix of the daughter rubidium atoms), then our calculated $KL_2L_3(^1D_2)$ transitions energies change to 11 442.1 eV and 11 435.1 eV for the EC decay and IC processes, respectively. In such a case the above seeming agreements turn to real ones. It should be, however, noted that the performed correction is approximate and its estimated accuracy reaches a few eV (depending on the accuracy of the quantities used). Nevertheless, the above findings demonstrate the possibilities of the MCDF calculations applied. In addition, their results indicate a significant role of the "atomic structure effect" increasing the KLL transition energies by 7 eV even for rubidium which, moreover, belongs to the "5$s$ elements".

On the other hand, the agreement between the measured and our calculated relative energies is unsatisfying. With the exception of the $KL_3L_3$ doublet, differences for the individual transitions are getting bigger and bigger as one moves away from the reference peak $KL_2L_3(^1D_2)$ (see the last column). As a result, the calculated spectrum is wider by 13.8(4) eV than the measured one (the opposite trend than in the case of the semi-empirical calculations [1]). At present, no explanation of this behavior is available.

As mentioned above, the influence of the atom environments on energies of the KLL Auger transitions in rubidium was investigated using one $^{83}$Sr source prepared by ion implantation into the platinum foil and three different $^{85}$Sr samples, namely $^{85}$Sr deposited by vacuum evaporation onto the carbon foil and implanted into both the platinum and carbon foils. Unfortunately, in the latter case we were able to measure only the $KL_{2,3}L_3$ line group (see Fig.4) due to the low $^{85}$Sr activity.

As can be seen from Table 1, the relative energies of the KLL transitions in Rb measured with the evaporated (Evap(C)) and implanted (Impl(Pt)) sources agree within one standard deviation with each other. Thus within the uncertainties quoted, no environmental effect on the relative transition energies is observed.

The absolute energy of the $KL_2L_3(^1D_2)$ transition measured with the evaporated source is higher by 1.2(19) eV than that one determined for the implanted (in Pt) source (see Table 1) but the difference does not exceeds the 1σ uncertainty limit. In Table 2, the energy shifts between the different $^{85}$Sr sources are shown. These were obtained from the relative positions of the $KL_2L_3(^1D_2)$ and $KL_3L_3(^3P_2)$ lines (presented in the measured spectra of the all three investigated sources) on the electron retarding voltage scale [9] that enables one to reach higher accuracy. As can be seen from the table (the last column), about four times greater reduction in the absolute energies of the investigated Auger lines was observed for the $^{85}$Sr atoms incorporated into the carbon host matrix compared to those implanted into the platinum host relative to the case of the $^{85}$Sr atoms deposited on the carbon surface. An explanation of this observation is not apparent because the source-host matrices are quite different and complex.

It is obvious from the above experimental data that the choice of the host material is crucial, e.g., for super stable calibration $^{83}$Rb/$^{83m}$Kr electron sources which will be prepared by ion implantation in the KATRIN neutrino mass experiment [8].

*3.2. Transition intensities*

The measured KLL transition intensities for rubidium relative to the total intensity of the KLL group are compared in Table 3 ("adopted values") with both the values [20] obtained from a fit to available experimental data and results of the relativistic calculations [3] in intermediate coupling with configuration interaction. There are also given the experimental data [17] for the nearest neighbor element Kr (Z=36) since the transition intensities vary slowly with Z. Very good agreement is found not only between experimental data for corresponding KLL transitions in Rb and Kr but also between our data and predicted values [3] with the exception of the $KL_1L_3(^3P_2)$ spectrum component. The fitted values [20] are also close to our intensities with the exception of the $KL_1L_3(^3P_2)$ and $KL_2L_3(^1D_2)$ lines. However, uncertainties of the data [20] are too great.

We also performed the MCDF calculations (see Appendix) of the KLL transition rates in Rb. No difference was found between the values obtained for the corresponding KLL transitions following the creation of initial vacancies by the electron capture decay of $^{85}$Sr and internal conversion in the daughter $^{85m}$Rb isotope. The results obtained are given in the last column of the table. Our calculated transition rates are close to those of the calculations [3] and agree with the measured values within 1σ with the exception of the $KL_1L_1(^1S_0)$ and $KL_1L_3(^3P_2)$ transitions.

The experimental value of 0.133(11) determined for the $KL_1L_2(^3P_0/^1P_1)$ transition intensity ratio agrees within 2σ with those obtained from the previous calculations [3] and ours but is higher by 9σ than the prediction of 0.031 given by the nonrelativistic calculations [5] accounting for the intermediate coupling and configuration interaction (see also Fig.1). Thus it is evident that the influence of the relativistic effects on the intensity distribution between the $^1P_1$ and $^3P_0$ components of the $KL_1L_2$ doublet is appreciable even at Z=37.

As regards environmental effects, it is seen from Table 3 that intensities of the $KL_1L_2(^1P_1)$ and $KL_1L_3(^3P_1)$ transitions measured with the $^{85}$Sr source prepared by ion implantation into the Pt foil are higher than those obtained with the evaporated source on the carbon backing. In the case of the latter transition, the intensity differs by up to 30%. This finding could indicate that the $L_2$ and $L_3$ natural level widths of $^{85}$Rb in the implanted $^{85}$Sr source are greater than those in the evaporated one. However, the intensities of the $KL_2L_2(^1S_0)$, $KL_2L_3(^1D_2)$, and $KL_3L_3(^3P_0)$ lines for the implanted source are lower (but still within the 1σ error limits).

### 3.3. Natural widths of the KLL Auger lines

Natural widths of the rubidium KLL Auger lines determined from the measured spectra are displayed in Table 4. They are compared with the estimated values obtained as a sum of the corresponding experimental rubidium atomic level widths [21-23] listed in Table 5. Generally, reasonable agreement is found between the measured and estimated values partly due to large uncertainties in some experimental data. However, most important is the detailed comparison of the KLL widths measured for the implanted source with those estimated using the experimental natural K and L subshell atomic level widths [21]. In both measurements (this one and [21]), the sources were prepared on the same mass separator by ion implantation into a high purity polycrystalline Pt foil at 30 keV and measured with the same electron spectrometer [9]. Thus identical or very close environments for the $^{83}$Rb (Ref. [21]) and $^{85}$Rb (this work) atoms can be expected. As can be seen from Table 4, good agreement is seen between the measured and estimated (the column A) natural KLL line widths with the exception of the $KL_1L_1$ transition. The estimated value in this case is substantially higher (by about 30%) than the measured one. But this fact is true also for the other estimated $KL_1L_1$ values presented in Table 4. Narrowing of the $KL_1L_1$ Auger line should imply its lower intensity. But the measured $KL_1L_1$ intensity for the implanted source is lower only by about 3% than the calculated one [3] and by about 6% than our value (see Table 3). So we have no explanation for this observation. If one compares the weighted means (w.m.) of the measured KLL line widths for the both sources with those (the column D) estimated with the use of the weighted means of the rubidium K- and L-subshell atomic level widths [21-23], the situation is almost the same.

Because of the complex structure of both the investigated line groups and the single spectrum line shapes as well as low intensities of so called "satellite lines" (weak lines) of the KLL spectrum, natural widths of the KLL transitions in rubidium were determined with relatively large uncertainties in our work. With the exception of the $KL_3L_3$ line, natural line widths obtained for the implanted and evaporated $^{85}$Sr sources agree with each other within 1σ. Thus it is evident that for the investigation of the environmental effects on the KLL Auger transition widths, a much higher experimental precision is required (i.e. much higher both the source intensity and the absolute instrumental resolution).

### 4. Conclusion

We performed extensive experimental investigation of the environmental effects on the KLL Auger spectrum of rubidium emitted in the radioactive decays of $^{83}$Sr and $^{85}$Sr isotopes incorporated in different host matrices. The results obtained demonstrate that, among others, the choice of the host material for super stable calibration $^{83}$Rb/$^{83m}$Kr electron sources which will be prepared by ion implantation in the KATRIN neutrino mass experiment plays an important role. Results of our "*ab initio*" multiconfiguration Dirac-Fock calculations of the rubidium KLL transition energies revealed a significant role of the "atomic structure effect" even for rubidium.

**Acknowledgement**

The work was partly supported by grants GACR P 203/12/1896, RFFI 13-02-00756, and the Australian Research Council (grant no. DP140103317).

**Table 1**
Energies (in eV) of the KLL Auger transitions (relative to the $KL_2L_3(^1D_2)$ one) in Rb following the electron capture decay of $^{85}$Sr evaporated on a polycrystalline carbon foil (Evap(C)) and implanted into a platinum backing (Impl(Pt)).

| Transition | Experiment (this work) Evap(C) | Experiment (this work) Impl(Pt) | Experiment (this work) Adopted[b] | Theory [1] | Rb (Th [1] - Exp) | Kr[a] (Th [1] - Exp) | Theory (this work) | Rb Th (t.w.) - Exp |
|---|---|---|---|---|---|---|---|---|
| $KL_1L_1(^1S_0)$ | − 451.9(4)[c] | − 452.0(4) | − 452.0(3) | − 448.2 | 3.8(3) | 3.7(6) | −465.9 | − 13.9(3) |
| $KL_1L_2(^1P_1)$ | − 255.0(2) | − 255.0(2) | − 255.0(2) | − 253.3 | 1.7(2) | 1.7(4) | −262.5 | −7.5(2) |
| $KL_1L_2(^3P_0)$ | − 227.9(13) | − 227.7(7) | − 227.7(6) | − 221.9 | 5.8(6) | 5.6(7) | −231.5 | − 3.8(6) |
| $KL_1L_3(^3P_1)$ | − 190.5(4) | − 190.3(3) | − 190.4(2) | − 186.7 | 3.7(2) | 2.9(4) | −195.1 | − 4.7(2) |
| $KL_1L_3(^3P_2)$ | − 167.3(6) | − 166.6(5) | − 166.9(4) | − 162.4 | 4.5(4) | 3.4(5) | −170.3 | − 3.4(4) |
| $KL_2L_2(^1S_0)$ | − 65.4(7) | − 66.3(6) | − 65.9(5) | − 62.4 | 3.5(5) | 2.8(5) | −65.4 | 0.5(5) |
| $KL_2L_3(^1D_2)$ | 11441.5(16)[d] |  |  | 11435.4[d] | − 6.1(16) |  | 11441.8 (EC)[e,f] |  |
|  |  | 11440.3(11)[d] |  |  | − 4.9(11) |  | 11434.8 (IC)[e,g] |  |
| $KL_3L_3(^3P_0)$ | + 47.5(7) | + 48.4(6) | + 48.0(5) | + 49.1 | 1.1(5) | − 0.1(4) | +49.4 | 1.4(5) |
| $KL_3L_3(^3P_2)$ | + 69.4(2) | + 69.4(2) | + 69.4(2) | + 67.7 | − 1.7(2) | − 2.1(3) | +69.3 | − 0.1(2) |

[a] Experimental data taken from Ref. [17].
[b] Values based on the weighted means of the corresponding relative energies obtained for the implanted and evaporated $^{85}$Sr sources.
[c] − 451.9(4) means − (451.9 ± 0.4)
[d] The energy referenced to the Fermi level.
[e] The energy referenced to the vacuum level.
[f] The energy calculated for the $KL_2L_3(^1D_2)$ transitions following the creation of the K vacancies by the EC decay of $^{85}$Sr.
[g] The energy calculated for the $KL_2L_3(^1D_2)$ transitions following the creation of the K vacancies by the internal conversion in $^{85}$Rb.

**Table 2**
The measured energy shifts (in eV) of the $KL_2L_3(^1D_2)$ and $KL_3L_3(^3P_2)$ transitions in $^{85}$Rb between the $^{85}$Sr source prepared by vacuum evaporation on a carbon backing (Evap(C)) and those prepared by ion implantation into the carbon (Impl(C)) and platinum (Impl(Pt)) foils.

| Source | $KL_2L_3(^1D_2)$ | $KL_3L_3(^3P_2)$ | w.m.[a] |
|---|---|---|---|
| Evap(C) | 0.0 | 0.0 | 0.0 |
| Impl(Pt) | - 0.8(2) | - 0.6(2) | - 0.7(1) |
| Impl(C) | - 2.7(2) | - 3.0(2) | - 2.9(1) |

[a] w.m. means the weighted mean.

**Table 3**
Relative intensities ($KL_iL_j/\Sigma KLL$, %) of the KLL Auger transitions in Rb following the electron capture decay of $^{85}$Sr evaporated on a polycrystalline carbon foil (Evap(C)) and implanted into a platinum backing (Impl(Pt))

| | Experiment | | | | | Theory | |
|---|---|---|---|---|---|---|---|
| | This work | | | Ref. [17] | Ref. [20][a] | Ref. [3][b] | This work |
| Transition | Evap(C) | Impl(Pt) | Adopted[c] | Kr(Z=36) | Kr(Z=36) | | |
| $KL_1L_1(^1S_0)$ | 7.3(3)[d] | 7.6(3) | 7.5(3) | 7.6(3) | 6.9(6) | 7.8 | 8.1 |
| $KL_1L_2(^1P_1)$ | 13.8(5) | 14.8(5) | 14.6(6) | 15.2(3) | 14.3(15) | 15.6 | 15.6 |
| $KL_1L_2(^3P_0)$ | 1.8(3) | 2.2(3) | 2.0(3) | 1.7(2) | 1.9(9) | 1.8 | 1.8 |
| $KL_1L_3(^3P_1)$ | 5.9(4) | 7.7(4) | 6.7(4) | 6.8(2) | 6.0(21) | 6.8 | 7.0 |
| $KL_1L_3(^3P_2)$ | 5.6(4) | 4.4(4) | 4.5(5) | 3.4(2) | 3.3(14) | 3.4 | 3.6 |
| $KL_2L_2(^1S_0)$ | 4.1(3) | 3.4(3) | 3.6(3) | 3.5(2) | 3.9(4) | 3.4 | 3.4 |
| $KL_2L_3(^1D_2)$ | 46.7(9) | 45.7(9) | 46.6(10) | 47.4(8) | 50.4(24) | 46.4 | 45.7 |
| $KL_3L_3(^3P_0)$ | 2.9(3) | 2.5(3) | 2.7(2) | 2.9(2) | 2.8(9) | 2.9 | 2.9 |
| $KL_3L_3(^3P_2)$ | 11.9(3) | 11.7(3) | 11.8(3) | 11.5(2) | 10.5(17) | 11.8 | 11.8 |
| | | | | | | | |
| $KL_1L_2(^3P_0/^1P_1)$ | 0.132(12) | 0.133(11) | 0.133(11) | 0.114(8) | 0.133(65) | 0.113 | 0.115 |

[a] A fit to experimental data. Z=36 is the closest elements for which the data are available.
[b] Interpolated values for Z=37.
[c] Values based on the weighted means of the corresponding relative intensities obtained for the implanted and evaporated $^{85}$Sr sources.
[d] 7.3(3) means 7.3 ± 0.3.

## Table 4
Natural widths $\Gamma_{KXY}$ (in eV) of the KLL Auger transitions in Rb following the electron capture decay of $^{85}$Sr evaporated on a polycrystalline carbon foil (Evap(C)) and implanted into the platinum backing (Impl(Pt)).

| Transition | Experiment (this work) | | | Estimated[a] | | | |
|---|---|---|---|---|---|---|---|
| | Evap(C) | Impl(Pt) | w.m.[b] | A[c] | B[d] | C[e] | D[f] |
| $KL_1L_1$ | 9.3(13)[g] | 8.2(10) | 8.6(8) | 10.4(4) | 10.6(12) | 11.8(5) | 10.9(3) |
| $KL_1L_2$ | 7.8(9) | 8.1(6) | 8.0(5) | 8.0(4) | 8.1(9) | 8.8(4) | 8.3(3) |
| $KL_1L_3$ | 8.1(19) | 7.3(8) | 7.4(8) | 8.0(5) | 8.1(9) | 8.7(3) | 8.2(3) |
| $KL_2L_2$ | 5.5(20) | 5.3(13) | 5.4(11) | 5.6(5) | 5.6(5) | 5.8(3) | 5.7(2) |
| $KL_2L_3$ | 4.8(2) | 5.1(2) | 5.0(1) | 5.6(5) | 5.6(5) | 5.7(3) | 5.6(2) |
| $KL_3L_3$ | 4.6(4) | 5.6(4) | 5.1(3) | 5.6(6) | 5.6(5) | 5.6(3) | 5.5(2) |

[a] $\Gamma_{KXY} = \Gamma_K + \Gamma_X + \Gamma_Y$ where $\Gamma_K$, $\Gamma_X$, and $\Gamma_Y$ are natural widths of the rubidium K and L-subshell atomic levels.
[b] w.m. means the weighted mean.
[c] Rubidium K and L subshell atomic level widths taken from Ref. [21].
[d] Rubidium K and L subshell atomic level widths taken from Ref. [22].
[e] Rubidium K and L subshell atomic level widths taken from Ref. [23].
[f] The weighted means of the rubidium K and L subshell atomic level widths [21-23] used.
[g] 9.3(13) means 9.3 ± 1.3.

## Table 5
Experimental rubidium K and L subshell atomic level widths used for the estimation of the rubidium KLL Auger line widths presented in Table 4.

| Subshell | Ref. [21] | Ref. [22] | Ref. [23] | w.m.[a] |
|---|---|---|---|---|
| K | 2.8(2)[b] | 3.0(2) | 3.0(2) | 2.9(1) |
| $L_1$ | 3.8(2) | 4.4(3) | 3.8(8) | 4.0(2) |
| $L_2$ | 1.4(3) | 1.4(1) | 1.3(3) | 1.4(1) |
| $L_3$ | 1.4(4) | 1.3(1) | 1.3(3) | 1.3(1) |

[a] w.m. means the weighted mean.
[b] 2.8(2) means 2.8 ± 0.2.

# APPENDIX

## Multiconfiguration Dirac-Fock Calculations of the KLL Auger Transitions in $^{85}$Rb Following the Electron Capture Decay of $^{85}$Sr

For an electron-capture-triggered K hole in $^{85}$Rb, the possible KLL-Auger decay processes are given by

$$
\begin{aligned}
^{85}Rb^*(1s^15s^2) &\rightarrow {}^{85}Rb^*(2s^05s^2) \quad + e_A^- \; (\text{KL}_1\text{L}_1) \\
&\rightarrow {}^{85}Rb^*(2s^12p^55s^2) + e_A^- \; (\text{KL}_1\text{L}_{2,3}) \\
&\rightarrow {}^{85}Rb^*(2p^45s^2) \quad\; + e_A^- \; (\text{KL}_{2,3}\text{L}_{2,3})
\end{aligned}
\quad (A1)
$$

where $e_A^-$ is a KLL Auger electron. Here the electron configuration is abbreviated to show only the electron vacancies and the valence configuration which remains that of Sr. The initial and final state wavefunctions were evaluated using the multiconfiguration Dirac-Fock (MCDF) method by applying the GRASP2K code [24] in the extended optimal level (EOL) scheme [25], where optimization is on a weighted sum of energies, together with the RECLI extension code [26]. Detailed descriptions of MCDF can be found elsewhere (see Ref. [27] and references therein). Only the main principles are reviewed here. In the MCDF method, the atomic state function (ASF) is represented as a superposition of $jj$-coupled configuration state functions (CSF) of the type

$$\Psi(vPJM_J) = \sum_{k=1}^{n_c} c_k \, \Phi(v_k PJM_J) \qquad (A2)$$

where $\Psi$ and $\Phi$ are, respectively, the ASF and CSF; $P$, $J$, and $MJ$ are the relevant quantum numbers: parity, total angular momentum, and the magnetic quantum number, respectively; $v$ represents the other quantum numbers that are necessary to describe the ASFs and CSFs, for example orbital occupancy and the angular-momentum coupling scheme. The summation in Eq (2) is up to $n_C$, the number of CSFs in the expansion and each CSF is built from antisymmetrized products of the one-electron Dirac orbitals, represented on a numerical mesh, for which initial estimates were obtained using the Thomas-Fermi model. In the relativistic self-consistent field procedure, both the expansion coefficients, $c_k$, and the radial parts of the Dirac orbitals, are optimized to find self-consistent solutions of the Dirac-Coulomb Hamiltonian

$$H_{DC} = \sum_i h_D(\mathbf{r_i}) + \sum_{i>j} \frac{1}{r_{ij}} \qquad (A3)$$

In this Hamiltonian $h_D(\mathbf{r})$ denotes the standard one-electron Dirac Hamiltonian in the field of the nucleus. The Breit interaction and the vacuum polarization corrections were included in subsequent relativistic configuration interaction (RCI) calculations by using the RELCI code [26].

The AUGER component [28] of the RATIP package [29] has been adopted to calculate the Auger transition energies and transition rates. In these calculations, a central-field approximation has been adopted to treat the interaction of the outgoing electron with the bound-state electron density. This results in continuum spinors which are obtained independently for each final state of the KLL Auger decay process within a spherical but level-dependent potential of the final ion (the so-called *optimal level* scheme). The AUGER calculations also incorporate the exchange interaction of the emitted electron with the bound electrons.

**Figure captions**

Fig. 1 The $KL_1L_2(^3P_0/^1P_1)$ Auger transition intensity ratio, as a function of the atomic number Z. Results of both relativistic [2,3] and nonrelativistic [5] calculations in intermediate coupling with configuration interaction (ICCI) are compared with experimental data (Ref. [4] and references therein). Experimental values obtained from fully or partly resolved $KL_1L_2$ doublets are denoted as "more reliable" and those from unresolved $KL_1L_2$ line as "less reliable". Our value for Z=37 is shown by full circle.

Fig. 2 The KLL Auger spectrum of Rb generated in the electron capture decay of radioactive $^{85}$Sr atoms implanted into a platinum foil. The spectrum was taken at an instrumental resolution of 7 eV and the 2 eV step in 13 sweeps. The exposition time per spectrum point in each sweep was 60 s. Results of the spectrum decomposition are shown by continuous lines.

Fig. 3 An example of the KLL Auger spectrum of Rb generated in the electron capture decay of $^{85}$Sr. The source used was prepared by thermal evaporation on a polycrystalline carbon foil. The spectrum was measured in eight sweeps with 60 s exposition time per spectrum point in each sweep and with 7 eV instrumental resolution and 2 eV step size. Continuous lines represent results of the spectrum decomposition into components.

Fig. 4 A fragment of the KLL Auger spectrum of Rb (the $KL_{2,3}L_3$ line group) generated in the electron capture decay of $^{85}$Sr implanted into a polycrystalline carbon foil. The spectrum was measured with 7 eV instrumental resolution and 2 eV step size in 20 sweeps with 60 s exposition per spectrum point in each sweep.

Fig. 5 An overview low-energy electron spectrum emitted in the $^{83}$Sr decay measured with 21 eV instrumental resolution and 7 eV step and with 30 s exposition time per spectrum point after four $^{83}$Sr half-lives from the source preparation. The spectrum was not corrected for the $^{83}$Sr decay as well as for the spectrometer transmission drop [9,10] with increasing electron retarding voltage. In the spectrum, electrons following the EC decay of $^{83}$Sr to $^{83}$Rb and $^{83}$Rb to $^{83}$Kr are seen. The insert shows, on an enlarged scale, a spectrum region including a part of the KLL Auger spectrum of Kr and the full KLL Auger spectrum of Rb.

Figure 1

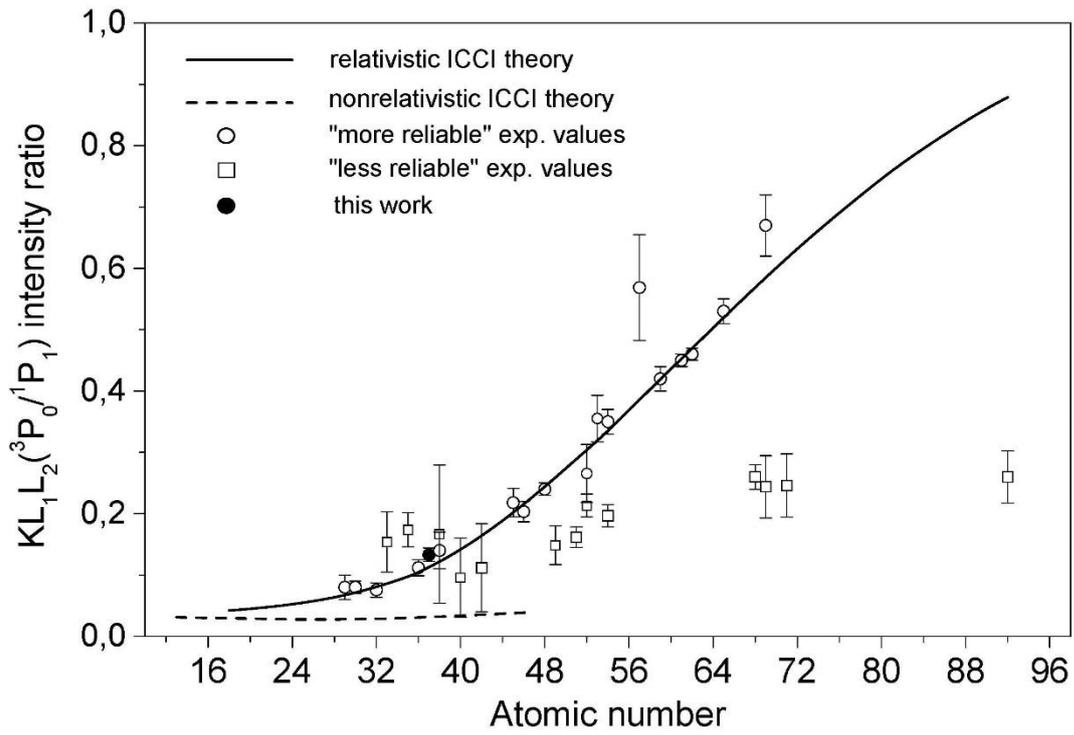

Figure 2

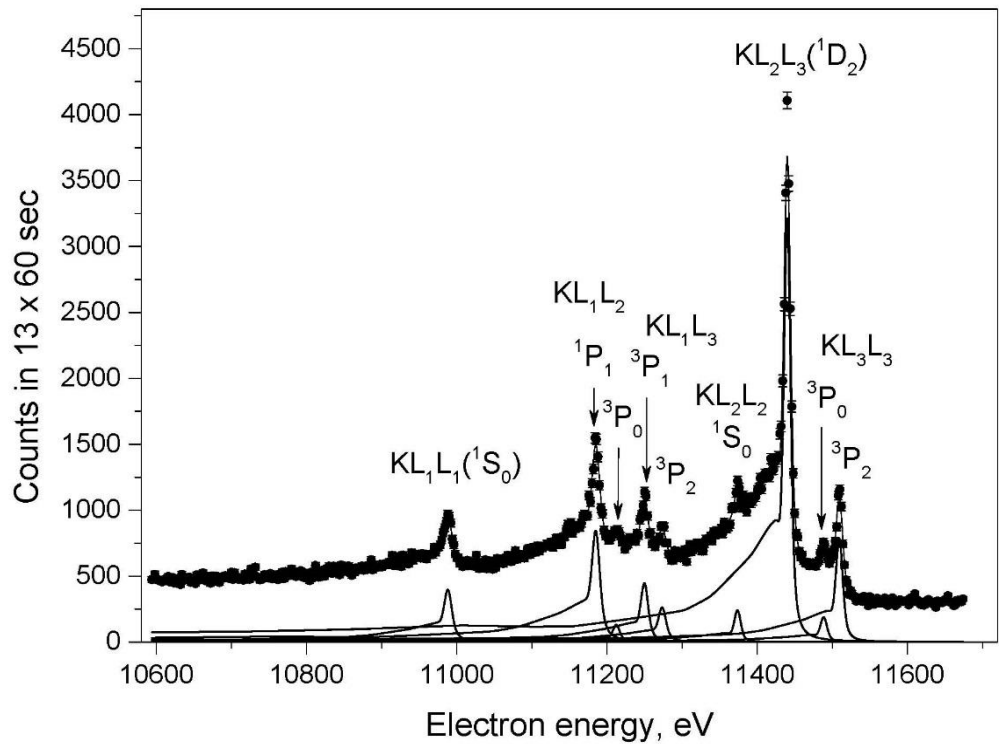

Figure 3

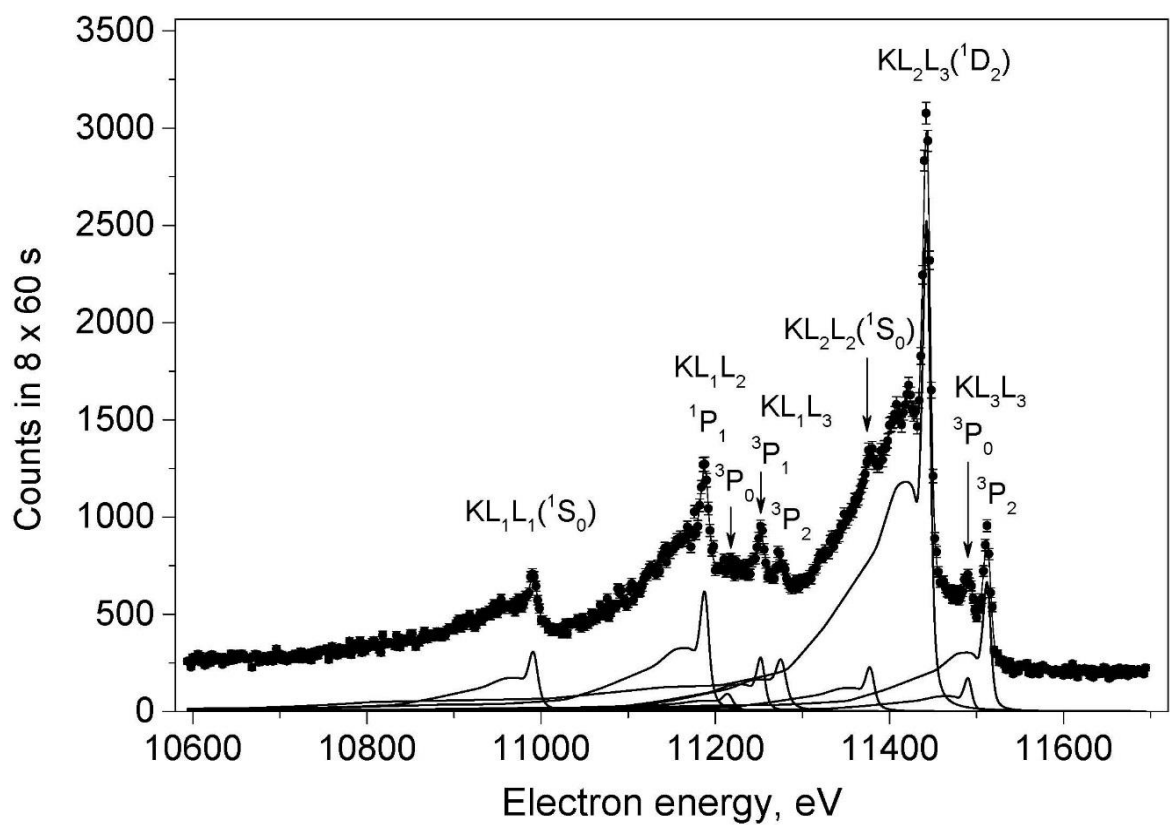

Figure 4

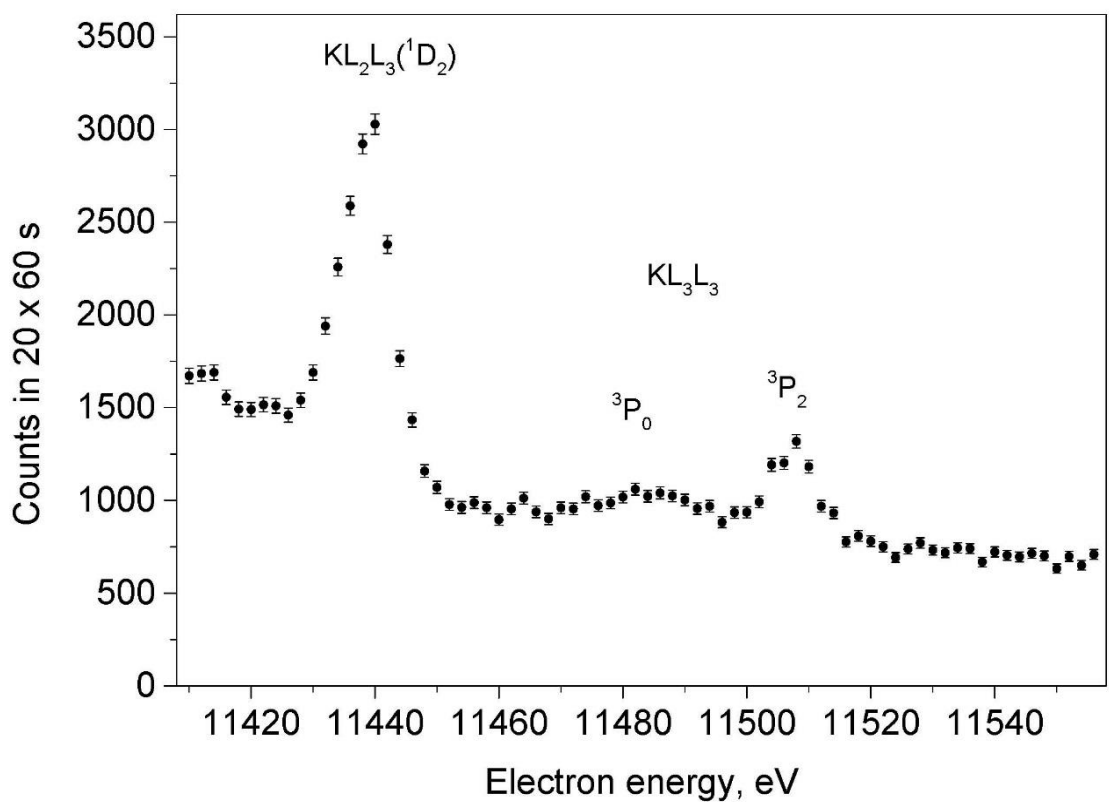

Figure 5

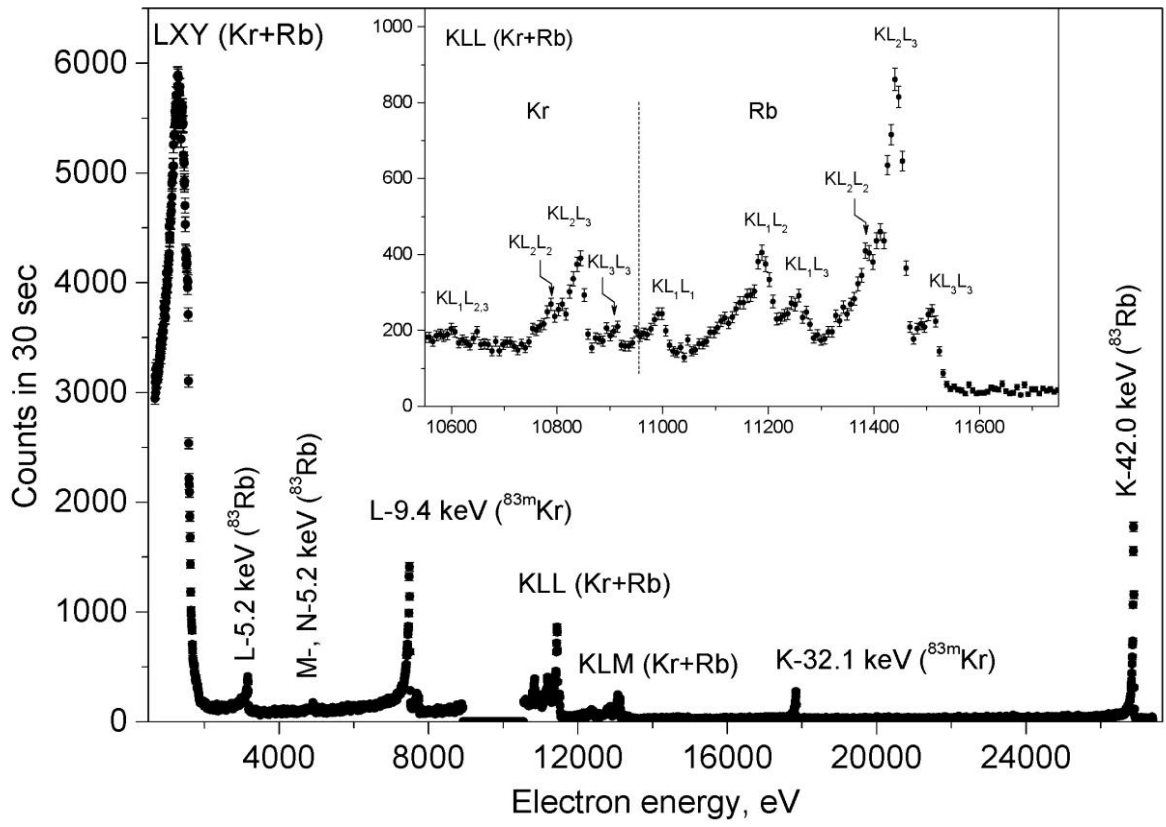